\documentclass[useAMS,usenatbib]{mn2e}
\newcommand{\Msun}{\mbox{$M_{\odot}$}}

\newcommand{\vvc}{$\Omega/\Omega_{\rm{crit,ZAMS}}$}
\usepackage{graphicx}
\usepackage{natbib}
\usepackage{amssymb}
\usepackage{hyperref}
\usepackage{amsmath}
\usepackage{lscape}
\usepackage{longtable}
\usepackage{caption}
\usepackage{xcolor}
\usepackage{aas_macros}
\hypersetup{colorlinks=true,citecolor=blue}

\definecolor{change}{RGB}{129,20,77}
\title[The enigma of NGC 2509]{Extended main sequence turnoffs in open clusters as seen by Gaia - II. The enigma of NGC~2509}
\author[M. de Juan Ovelar et al.]{M. de Juan Ovelar$^1$\thanks{m.dejuanovelar@ljmu.ac.uk}, S. Gossage$^{2}$, S. Kamann$^{1}$, N. Bastian$^{1}$, C. Usher$^{1}$,  
\newauthor I. Cabrera-Ziri$^{2}$\thanks{Hubble fellow}, A. Dotter$^{2}$, C. Conroy$^{2}$ and C. Lardo$^{3}$\\
$^{1}$ Astrophysics Research Institute, Liverpool John Moores University, 146 Brownlow Hill, Liverpool L3 5RF, UK\\
$^{2}$ Harvard-Smithsonian Center for Astrophysics, 60 Garden Street, Cambridge, MA 02138, USA\\
$^{3}$ Laboratoire d’astrophysique, Ecole Polytechnique Fédérale de Lausanne (EPFL), Observatoire de Sauverny, CH-1290 Versoix, Switzerland
}
\begin{document}

\date{Accepted 2019 November 03. Received 2019 October 16; in original form 2019 August 13 }

\pagerange{\pageref{firstpage}--\pageref{lastpage}} \pubyear{2019}

\maketitle

\label{firstpage}

\begin{abstract}

We investigate the morphology of the colour-magnitude diagram (CMD) of the open cluster NGC~2509 in comparison with other Galactic open clusters of similar age using Gaia photometry. At $\sim~900~\rm{Myr}$ Galactic open clusters in our sample all show an extended main sequence turn off (eMSTO) with the exception of NGC~2509, which presents an exceptionally narrow CMD. 
Our analysis of the Gaia data rules out differential extinction, stellar density, and binaries as a cause for the singular MSTO morphology in this cluster.
{We interpret this feature as a consequence of the stellar rotation distribution within the cluster and present the analysis with MIST stellar evolution models that include the effect of stellar rotation on which we based our conclusion. In particular, these models point to an unusually narrow range of stellar rotation rates (\vvc$ = [0.4,0.6]$) within the cluster as the cause of this singular feature in the CMD of NGC~2509. Interestingly, models that do not include rotation are not as good at reproducing the morphology of the observed CMD in this cluster.}  

\end{abstract}


\section{Introduction}
\label{sec:intro}

%%%% FIGURE %%%
\begin{figure*}
\centerline{\includegraphics[scale=0.65]{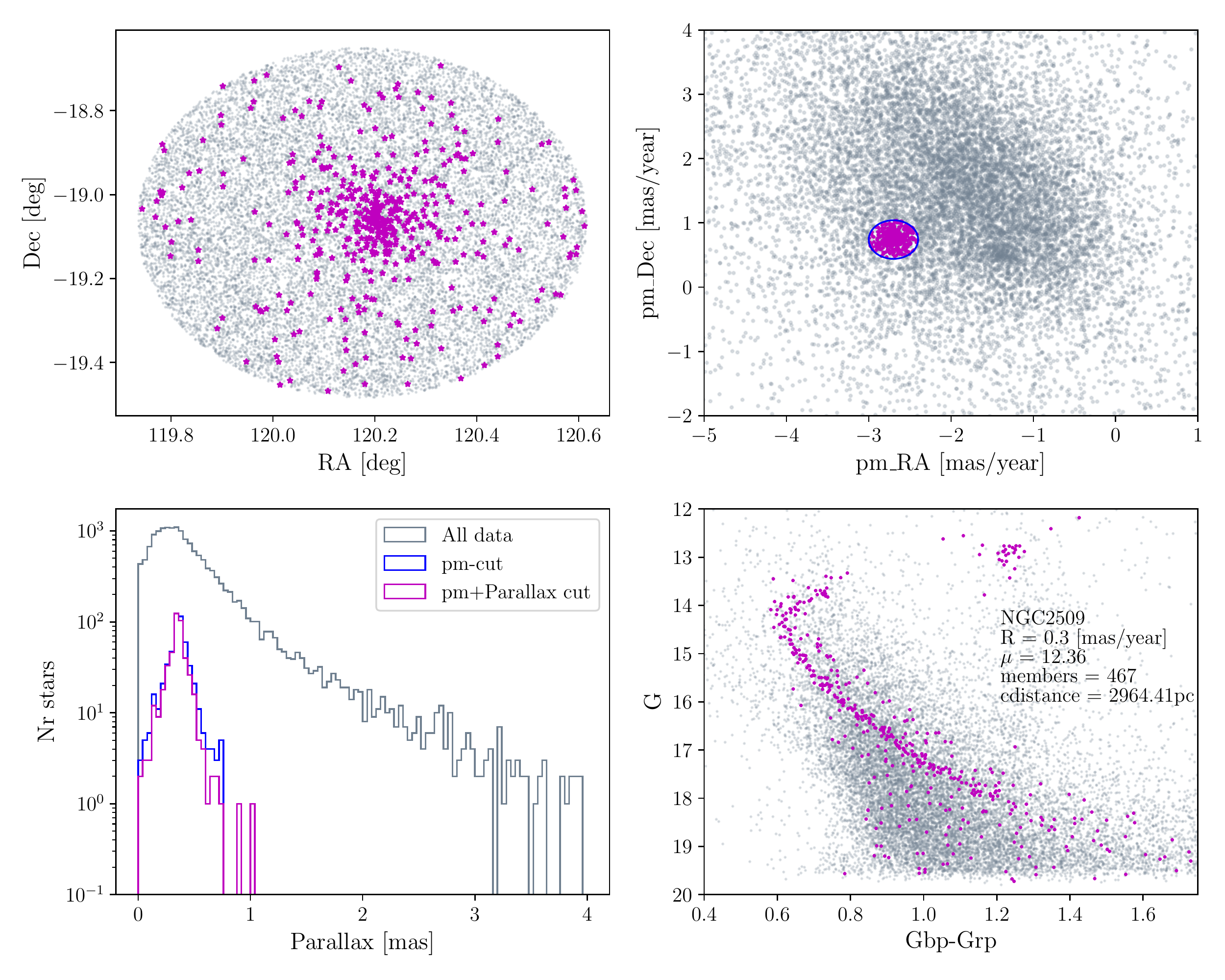}}
\caption{{Upper left:}  Spatial distribution of the stars in the Gaia catalogue within $10~\rm{arcmin}$ of NGC~2509's centre (hereafter referred to as field of view). Cluster members are highlighted as magenta-filled stars. {Upper right:} Proper motions of all stars, with cluster members shown as magenta-filled dots. The blue circle marks the area occupied by the selected members in proper motion space. {Bottom left:} Parallax distribution of all stars in the field of view (gray-solid line), stars within the proper-motion area we defined for the cluster (blue-solid line) and selected members based on the proper motions and parallax cuts applied (see Section~\ref{sec:membership}). {Bottom right:} Colour-Magnitude diagram of all stars in the field with cluster members shown again as magenta-filled dots.}
\label{fig:selection}      
\end{figure*}
%%%% FIGURE %%%

Originally discovered in massive intermediate age stellar clusters in the Large Magellanic Clouds  \citep{mackeynielsen07} extended main sequence turn-offs (eMSTOs) are now known to be a common feature of clusters with ages between $10-20$~Myr \citep{li17,beasor19} and $\sim2$~Gyr \citep[e.g.~][]{martocchia18} and masses as low as a few thousand solar masses \citep[e.g.~][]{piattibastian16}.  Additionally, with proper motion cleaned high precision photometry available with Gaia, eMSTOs have been seen in an increasing sample of Galactic open clusters as well \citep{bastian18,marino18b,cordoni18}.

The ubiquity of the eMSTO feature, as well as the observed relation between the age of the cluster and the extent of the MSTO width \citep{niederhofer15} argue for a stellar evolutionary effect as the cause of the feature.  One such effect could be stellar rotation, where rotating stars have their position in a colour-magnitude diagram (CMD) shifted due to changes in both their internal chemical and hydrostatic equilibrium structures \citep{bastiandemink09}.  Indeed, the position of a star within the CMD on the MSTO has been found to be correlated with the measured rotational velocity ($Vsini$) in a number of clusters \citep{dupree17,kamann18b,bastian18,marino18b}.

As part of a large study on the extended MSTO phenomenon, we have compiled a list of $\sim50$ open clusters with precise parallax determination, low differential extinction, and with ages that span from $\sim30~\rm{Myr}$ to $\sim4~\rm{Gyr}$.  The full sample and analysis will be presented in a future work (de Juan Ovelar et al. in preparation).  As part of this analysis, we found a specific cluster, NGC~2509, which deserves special attention, given its apparent lack of an eMSTO at an age ($\sim900~\rm{Myr}$) where all other coeval clusters in the sample clearly show it.  In the present study we analyse the CMD of NGC~2509 and compare it with other clusters in our sample with similar ages using photometry obtained with Gaia. In particular, we compare the eMSTO properties of NGC~2509 to NGC~1817, NGC~2360, NGC~2818, NGC~5822, NGC~7789 and Melotte~71, focusing on its lack of extended MSTO in comparison with the others. We argue that the particular morphology of NGC~2509's MSTO is related to the particulars of its member's rotation distribution.

In Section \ref{sec:methods} we present the data and the process followed to select members of the clusters, and to estimate their extinction, ages and age-spread. In Section \ref{sec:analysis} we analyse NGC~2509's CMD in comparison with the other clusters, and in Section \ref{sec:discussion} we discuss the possible explanations of the differences we find. Finally, we present our conclusions in Section \ref{sec:conclusions}.

%%%%%%%%%%%%%%%%%%%%%%%%%%%%%%%%%%%%
\section{Data and methodology}
\label{sec:methods}

%=====================================
\subsection{Membership selection} \label{sec:membership}

We use Gaia DR2 data of the following clusters: NGC~1817, NGC~2360, NGC~2509, NGC~2818, NGC~5822, NGC~7789, and Melotte~71.  These clusters are those in our sample with clean CMDs and ages within $200~\rm{Myr}$ of $\sim900~\rm{Myr}$, which is our estimated age for NGC~2509. We downloaded Gaia DR2 astrometry, proper motions, photometry, and parallaxes of stars within $10~\rm{arcmin}$ of the centre of each cluster. 

The cluster member selection is done systematically as follows. First we clean the downloaded catalog by rejecting sources with an error in proper motion larger than $0.5~\rm{mas/year}$. We then determine the cluster centre either by means of a 2D-histogram of the remaining stars in proper motion space, or by directly inspecting the proper motion diagram. Then, we select stars within a certain radius of this centre in proper motion space, which is different for each cluster, and build a histogram of their parallaxes. Any star with parallax value within $2\sigma_t $ of the maximum of the parallax distribution, with $\sigma_t $ being each individual star's error in parallax, is considered a candidate member. {At this point, we refine the selection applying a further 3D--membership probability cut in proper motion and parallax space following the method outlined in \citet{kamann14} and assuming a velocity dispersion of $2~\rm{km/s}$ independently of the cluster.} 

Figure~\ref{fig:selection} shows images of spatial, proper motion, parallax and color-magnitude distributions for the cluster NGC~2509 used to select the cluster members (top-left, top-right, bottom-left and bottom-right panels, respectively). Once we have a catalog of members for each cluster, we obtain an estimate of the cluster distance $d$ by computing the median value of the parallax of all members above a G magnitude of 17. {We then obtain the distance modulus ($\mu$) of each cluster as $\mu=5\times\log(d) - 5$. In order to investigate the effect that the zero point offset issues in the Gaia parallaxes \citep[e.g.][]{lindegren18} have in our $\mu$ values, we have recomputed all distance modulii in our sample subtracting 0.08 mas from the median parallax measured for each cluster, which is the maximum average global offset we found in the literature \citep[see][]{choi18}. The average variation in the values of $\mu$ for our sample is of a 3.3\%, with a maximum of 5.8\% for the cluster NGC~2818. Keeping in mind that we assumed the worst case scenario for a global offset in parallax, we can assume that the impact of this issue in our particular sample is very low.}

%=====================================
\subsection{Estimating extinction and age}\label{subsec:isochfit}

%%%% FIGURE %%%
\begin{figure}
\centerline{\includegraphics[scale=0.55]{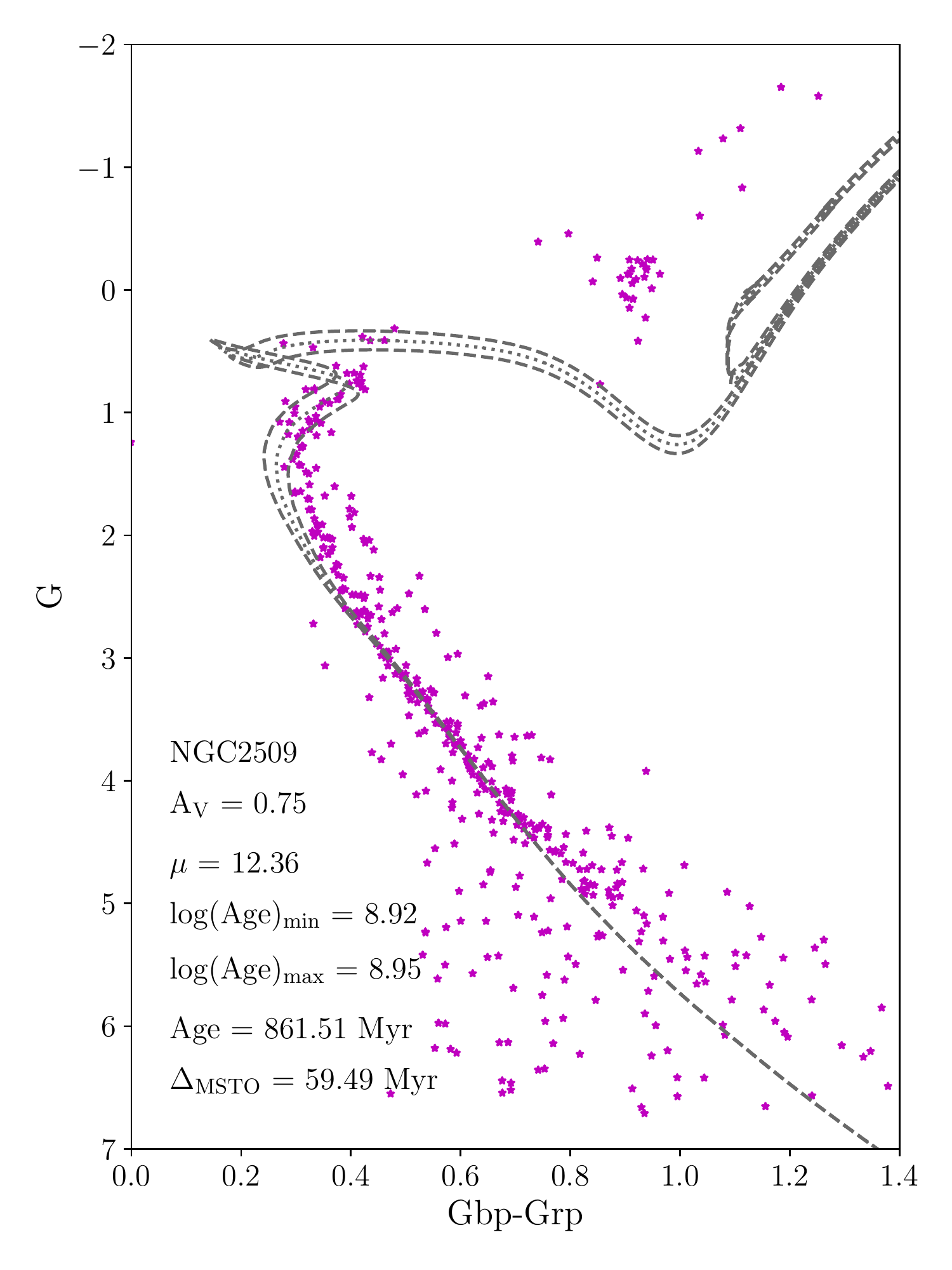}}
\caption{Colour-Magnitude diagram of NGC2509's cluster members shown as magenta-filled stars. Three non-rotating MIST isochrones are shown together with the data. The two gray-dashed lines correspond to the isochrones delimiting the spread of the main sequence turn-off in NGC2509 ($\Delta_{\rm{MSTO}}$). The gray-dotted line is the nearest isochrone to our estimated age.}
\label{fig:agefit}
\end{figure} 
%%%% FIGURE %%%

In order to estimate the extinction (A$_{\rm V}$) and age of the clusters, we first {fit} non-rotating Mesa Isochrones and Stellar Tracs (MIST) isochrones \citep{dotter16,choi16} with a range of ages (from $10~\rm{Myr}$ to $10~\rm{Gyr}$) to the main sequence (MS) region of the CMD using A$_{\rm V}$ as a free parameter. We then proceed to find the two isochrones {delimiting} the MSTO's blue and red edges ($\log (\rm{Age})_{\rm{min}}$ and $\log (\rm{Age})_{\rm{max}}$, respectively), which we then use to define the spread of the MSTO ($\Delta_{\rm{MSTO}}$) as the difference (in Myr) between the two isochrones.  We finally compute the age as the middle point between these two {delimiting} isochrones (again in Myr). The values obtained for the parameters of each cluster are shown in Table~\ref{tab:parameters}.  Figure~\ref{fig:agefit} shows our age fit for NGC~2509 as an example.  Note that {all of the isochrone fitting is done visually} and so, we expect some differences between the values found in this study and those across the literature.  {However, it is important to note too} that the absolute values we find {through this method} are not very relevant to the study since our analysis and statistics are done in relative terms, i.e.~comparing cluster's CMD morphologies directly (see next section). 

{Following this isochrone fitting procedure, we find an age for NGC~2509 of $860~\rm{Myr}$, and a MSTO spread of $60~\rm{Myr}$. Note that his value for the MSTO spread is very low compared to the other clusters in our sample which, with ages $900\pm200~\rm{Myr}$, all present spreads of at least $200~\rm{Myr}$ (see Table~\ref{tab:parameters}). In particular, the two clusters closest in age to NGC~2509, i.e.~MELOTTE~71 ($940~\rm{Myr}$) and NGC~5822 ($810~\rm{Myr}$), have spreads of $210~\rm{Myr}$ and $202~\rm{Myr}$, respectively.}

\setlength\tabcolsep{3.5pt}
\begin{table}
	% Stretch the rows a bit to give more space
	\renewcommand{\arraystretch}{1.3}
	\renewcommand{\captionfont}{\scshape}
	% Changes table font (not caption! see above)
	\small	
	\begin{center}
		\caption{Estimated parameters for our cluster sample}
		\label{tab:parameters}
		\begin{tabular}{l|c|c|c|c|c|c|c}
			\hline 	
			{Cluster} 		& {$\mu$} 		& {$A_{\rm V}$} 	&  {$\rm{Age}$}	& $\Delta_{\rm{MSTO}}$	& {$N^{\star}_{\rm{MS}}$}	& {$f_{\rm{bin}}$}	& {$r_{\rm{MSTO/MS}}$}\\
			\hline 
			NGC~1817	& $11.37$		& $1.1$			& $720$		& $230$ 				& $140$				& $0.17$			& $3.36$\\
			NGC~2360	& $10.23$		& $0.5$			& $960$		& $370$ 				& $101$				& $0.21$			& $3.79$\\
			NGC~2509	& $12.36$		& $0.75$			& $860$		& $60$ 				& $116$				& $0.17$			& $1.38$\\
			NGC~2818	& $12.82$		& $0.9$			& $720$		& $230$				& $179$				& $0.19$			& $3.27$\\
			NGC~5822	& $9.63$		& $0.6$			& $810$		& $202$ 				& $91$				& $0.18$			& $4.18$\\
			NGC~7789	& $11.76$		& $1.3$			& $1070$		& $390$ 				& $736$				& $0.19$			& $2.41$\\		
			MELOTTE~71	& $11.81$		& $0.65$			& $940$		& $210$				& $184$				& $0.16$			& $3.77$\\
			\hline
		\end{tabular}
	\end{center}
	%[1.5] %You can adjust how far below the table the text should appear
  \scriptsize{Cluster, distance modulus, extinction, age (Myr), main sequence turn-off spread (Myr), number of stars in main sequence subsample, binary fraction in (observed, see text) and ratio of standard deviation of distribution of distances to line A in MSTO subsample, over standard deviation of distribution of distances to line B in MS subsample (see text and Figure~\ref{fig:distributions}).}\\
\end{table}

%%%%%%%%%%%%%%%%%%%%%%%%%%%%%%%%%%%%
\section{Analysis}
 \label{sec:analysis}
 
 %%%% FIGURE %%%
\begin{figure*}
\centerline{\includegraphics[scale=0.65]{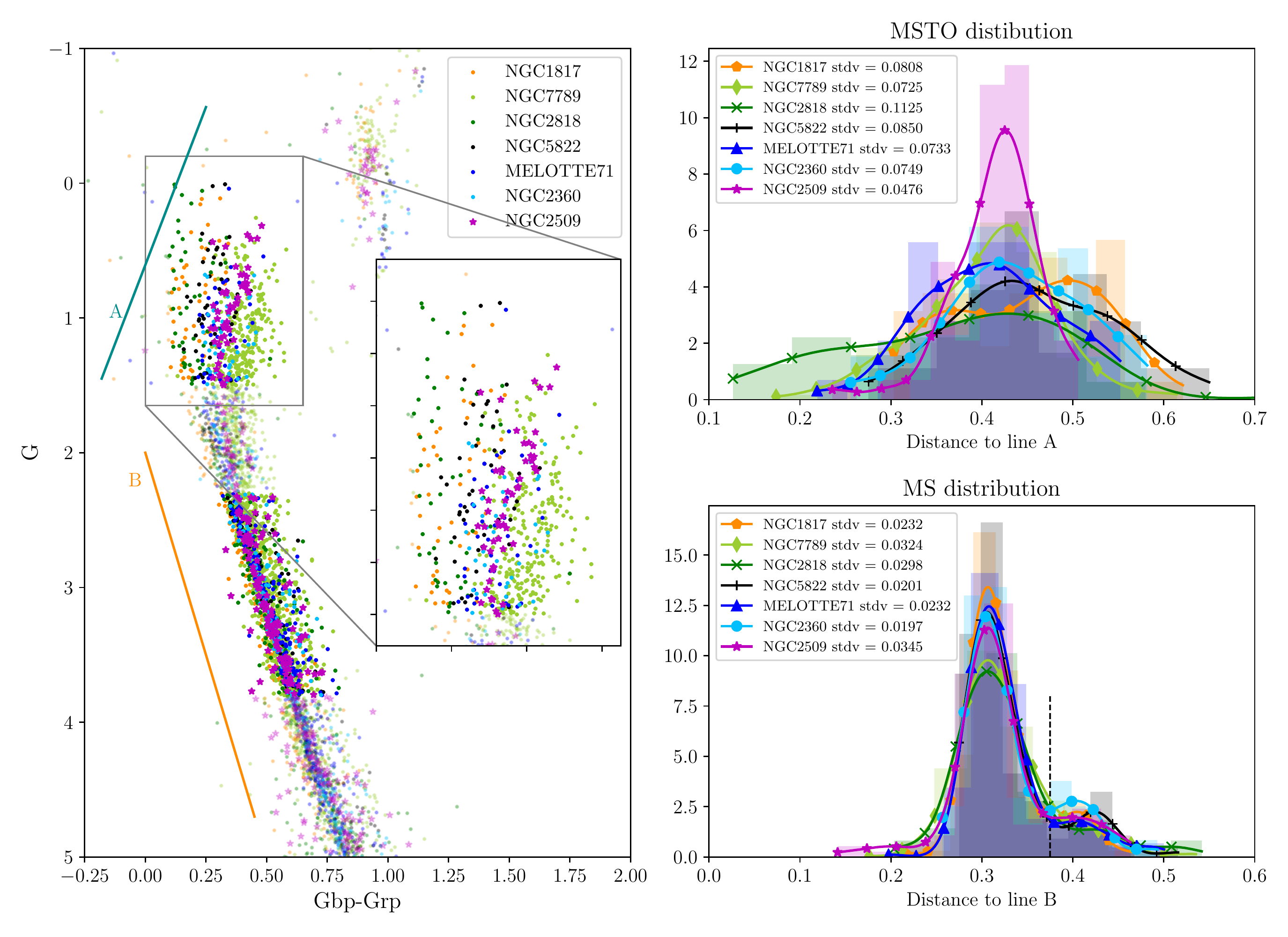}}
\caption{{Left panel:} Colour-Magnitude diagram of all clusters included in this study shown as colour-coded filled dots except for NGC~2509 which is shown in magenta-filled stars. The blue and orange solid lines designated as A and B are the lines used for the analysis of the main sequence (MS) and main sequence turn-off (MSTO) regions of each cluster's diagram (see Section~\ref{sec:analysis}). A detail of the MSTO region of the diagram is shown in the zoomed-in inset. {Upper right:} Histograms and Kernel Density Estimates (KDEs) computed for the distributions of distances of stars in the MSTO regions of each cluster to the line A. {Lower right:} Same as the upper panel but computed for the distributions of distances in the MS region to line B. {Note that solid-fill symbols highlight the members selected to represent these two regions}. The black-dashed vertical line marks the separation we define between stars in the MS (to the left of the line), and stars in the binary sequence of the clusters. The legend in each of these two right panels shows the computed standard deviations of the distributions, which, in the case of the MS region (lower right panel) excludes all stars in the binary sequence, i.e.~those to the right of the vertical black-dashed line.}
\label{fig:distributions}
\end{figure*}
%%%% FIGURE %%%

Because the ages and age spreads we obtain with the above procedure are dependent on a {visual} fitting of isochrones, we choose to compare the CMDs between clusters directly.  Figure \ref{fig:distributions} shows the result of this procedure for all clusters included in this study, with the left panel showing the CMDs of all clusters, once corrected for extinction and distance using A$_{\rm V}$ and the distance modulus ($\mu$), overplotted in different colours.  We then select cluster members in the MSTO by means of a cut in their color and magnitude.  In particular, we select members of all clusters with colour (G$_{\rm{bp}}$-G$_{\rm{rp}}$) and magnitude (G) values between $[0.1-0.6]$ and $[1.5-0.]$, respectively.  Then we compute the distances, in CMD space, of each member in this region to an arbitrary line that we fix for all clusters (blue line A in left panel of Fig.~\ref{fig:distributions}).  In this way, we are able to build a distribution of distances that gives us an idea of the spread of the MSTO sub-sample and allows us to compare the morphology of the MSTO of each cluster directly.  The normalised histograms and Kernel Density Estimates of these distribution are shown in the upper right panel of Figure~\ref{fig:distributions} for all clusters.  The histograms, shown as filled-bars, are computed using 10 bins for all clusters.  The KDEs, shown as solid lines, are computed using Gaussian kernels and Scott's ``rule of thumb" bandwidth selection method.  Note that the histograms and KDEs are computed independently from each other.  Due to the small differences in age between clusters, the distributions are slightly misaligned, so we correct for that aligning their peaks (or centres, in the case of bimodal distributions) with the peak in the MSTO distribution of NGC~2509.  This operation is just for visualisation purposes and does not affect the shape of the distribution.  In this panel, we can already see that the MSTO of NGC 2509 (solid magenta line) is significantly narrower than the MSTOs of the other clusters in the sample. 

In order to verify that the differences in the morphology of these MSTO distributions are not caused by e.g.~differential extinction that could be broadening the CMD of each cluster differently, we repeat this procedure on stars in the MS region of the CMD, selecting those with colour and magnitude values between $[0.3-0.8]$ and $[3.8-2.3]$, respectively, and defining a new, fixed, line in that region (orange line B in left panel of Fig.~\ref{fig:distributions}) and compute the distance.  We build the corresponding histograms and KDEs in the same way we did for the MSTO region.  The result of this analysis is shown in the lower-right panel of Fig.~\ref{fig:distributions}, where it is clear that the distributions, their widths in particular, are remarkably similar, ruling out the possibility of external factors (i.e.~differential extinction) being the cause of the differences seen in the MSTO region. In order to quantify this difference we compute standard deviations of all distributions of distances. The ratio between the MSTO distribution's standard deviation to the MS one gives an idea of how much more the MSTO is broadened with respect to the MS. This ratio for NGC~2509 is of $\sim1.4$ while the minimum ratio found in the rest of the sample clusters is $\sim2.4$ for NGC~7789, with a maximum of $\sim4.2$ for NGC~5822. This gives an idea of how different NGC~2509 is from the rest of the sample.  Note that we excluded what we identified as the binary sequence of our clusters \footnote{We define as the binary sequence distribution, the part of the distances distribution in the MS region to the right of the black-dashed vertical line in the lower-right panel of Figure~\ref{fig:distributions}} from the standard deviation calculation in the MS region. The effect of binaries in the broadening of the MSTO is not significant (see Section~\ref{sec:binaries}) so in this region we include the full range of the distribution.

Finally, in order to rule out differential extinction as a MSTO broadening factor in particular, we performed an experiment to measure its effect. We created synthetic clusters using the same non-rotating and solar metallicity MIST isochrones with no differential extinction we use for the age fitting procedure. We then add a range of values for differential extinction ($0.0$ to $0.3$) and we run our method on them. We then measure standard deviations in the same way as we did for our observed clusters and compute the ratios in MSTO standard deviation to MS one for the distribution of distances in the two regions of each cluster. We found a maximum ratio of $1.8$ in the worst differential extinction case ($0.3$). Note that these synthetic clusters do not have binaries. The ratios obtained for our observed clusters are much larger, which effectively rules out differential extinction as a possible cause for the differences we are finding between NGC~2509 and the rest of clusters in our sample.

%%%%%%%%%%%%%%%%%%%%%%%%%%%%%%%%%%%%
\section{Discussion}
\label{sec:discussion}

We have shown that the open cluster, NGC~2509, has a much narrower MSTO than all other clusters in our survey with similar ages.  The question, naturally, becomes why is this the case?  In order to answer this question we have looked at a number of things that could affect the spread in the MSTO region of the CMD in these clusters.

%======================
\subsection{Stellar density}
\begin{figure}
\centerline{\includegraphics[scale=0.5]{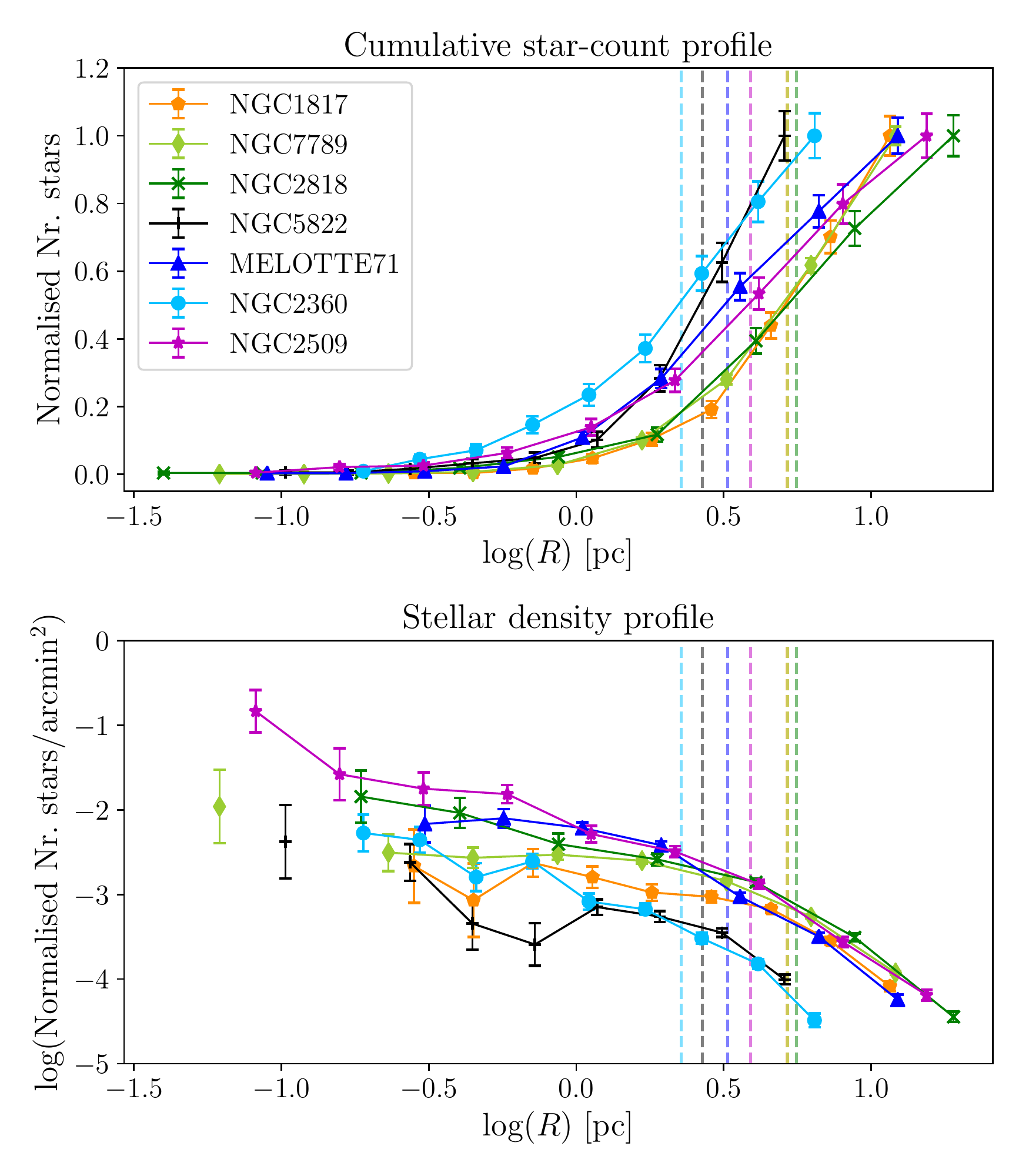}}
\caption{{Upper panel:} Cumulative star-count profile of the cluster members colour-coded for each cluster. The vertical-dashed lines mark the half-light radius of each cluster. {Lower panel:} stellar density profiles of the clusters}
\label{fig:radialprofiles}
\end{figure}

One potential cause for the differences between NGC~2509 and the other clusters could be if the clusters' stellar densities are significantly different (as this may affect the rotational distribution of stars).  Figure~\ref{fig:radialprofiles} shows the radial cumulative star-count and stellar density profiles (top and bottom panels, respectively) of all clusters included in this study. We can see here that NGC~2509 does not stand out as particularly special in terms of density. 

%%%% FIGURE %%%
%%%% FIGURE %%%

In order to further confirm this, we estimate the relative densities of all clusters in our samples using the number of stars used to fit the width of the MS region ($N^{\star}_{\rm{MS}}$ in Table~\ref{tab:parameters}) as a proxy for mass, and our estimates for the half-light radius of each cluster (dashed coloured lines in Fig.~\ref{fig:radialprofiles}).  Our choice of method to approximate the mass of the clusters is justified by the fact that the MS region is complete for all clusters and also has relatively small errors (on proper motions and parallax) so that background contamination is minimised. With these estimates for the relative density of the clusters we find that all clusters in our sample have similar densities within a factor of $5$. This rules out density as a cause for the lack of eMSTO in NGC~2509.  So next, we look at the rotation distribution within the cluster.

%======================
\subsection{Stellar rotation distribution}

\begin{figure}
\centerline{\includegraphics[scale=0.3]{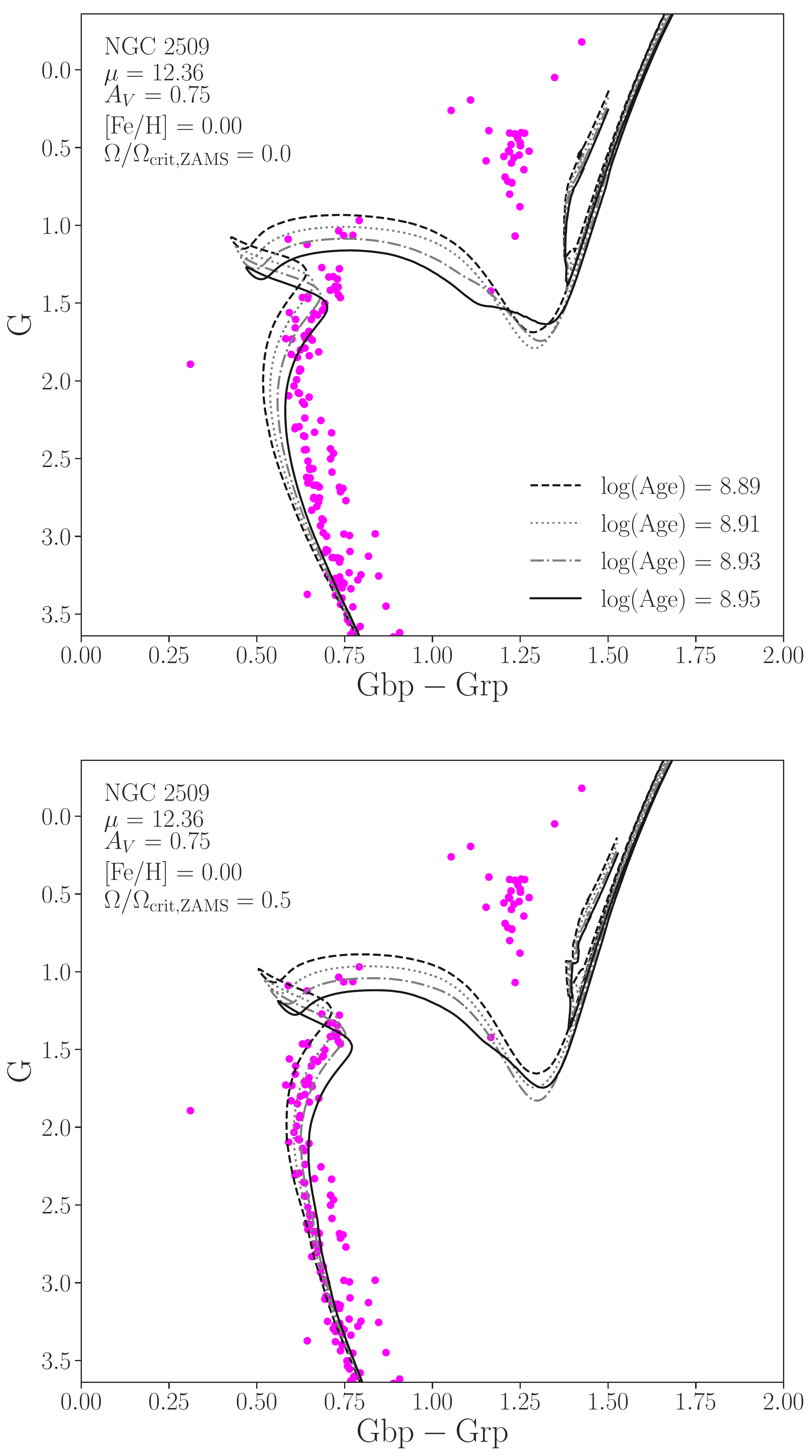}}
\caption{The observed colour magnitude diagram of NGC 2509 members with overplotted non-rotating (upper panel) and rotating(lower panel) isochrones for a range of ages from $\log(\rm{Age}) = 8.89$ to $8.95$.  The rotation rate assumed in the lower panel is our best fit value of \vvc $= 0.5$.  Note that the rotating isochrones match the morphology of the upper main sequence better than the non-rotating ones.}
\label{fig:varyage}
\end{figure}

In studies of the eMSTO phenomenon, the case has been made for a distribution of rotation rates as the cause behind the phenomenon \citep[e.g.][]{bastiandemink09, niederhofer15, brandthuang15b,  kamann18b, marino18b}. {Stellar rotation affects both the luminosity and temperature of a star, effectively changing its position in a CMD. On top of convective chemical mixing, rotating stars have rotationally induced mixing, which increases their luminosity and also their lifetimes \citep{maeder09}. Moreover, rotating stars have their spherical symmetry broken due to the centrifugal forces that now act on the physical structure. These introduce a latitudinal dependence on the effective surface gravity, which is directly linked to the effective surface temperature of the star. This means that the measured colour of a rotating star changes with the stellar inclination angle. Averaged over the surface, the temperature of a rotating star is lower than that of a non-rotating star with same characteristics. This effect is thus referred to as \textit{gravity darkening} \citep{vonzeipel1924}. If a stellar population presents a distribution of stellar rotation rates and inclinations within its members, the aforementioned effects can cause a spread of the MSTO region of the CMD which, as mentioned in the introduction, has been observationally found to be true in several clusters already \citep[e.g.][]{bastian18}. In this context, the lack of an eMSTO in NGC2509 may be related to the characteristics of the stellar rotation distribution of the cluster. In order to investigate this hypothesis, we present here some modelling efforts done with the MIST stellar evolution framework \citep{choi16,dotter16,choi18,gossage18}, which includes the effects of rotation.}

{We began with a qualitative analysis of the stellar rotation distribution in NGC~2509. Fixing metallicity, $\mu$, and extinction, we {visually} checked how well different non-rotating and rotating MIST isochrones fit the data at different ages. The results are shown in Figure \ref{fig:varyage}. In the top panel it is apparent that the non-rotating isochrones are always too blue near the MSTO at the fixed cluster parameters, regardless of age. We varied the rotation rate of the models until we found a visual fit to the data at \vvc=0.5, where \vvc~is the ratio of angular velocity at the zero age main sequence (ZAMS), divided by the critical angular velocity of the star  \citep[see][for details]{maeder09}. At this rotation rate, the blueward bend along the upper MS (lower MSTO) in the isochrones becomes less pronounced, due to the reddening effect of gravity darkening, providing a better qualitative fit to the data. Through this initial analysis, we thus suggest that NGC~2509 might be lacking a broad range of stellar rotation rates (slow rotators in particular), and seems to be primarily comprised of stars rotating near \vvc=0.5. Interestingly, this area in the blueward bend along the MS/MSTO where the non-rotating MIST isochrones miss the data in NGC~2509, seems to be well populated in the other clusters in our sample, which would fit in the hypothesis of the MSTO spreads in these clusters as being caused by a wide range of stellar rotational velocities present, with both slow and fast rotators within the population.} 

{There are, however, some important caveats of Figure \ref{fig:varyage} that ought to be considered. Regarding gravity darkening, the effect is present, but it is important to note in this figure that the plotted MIST rotating isochrones have luminosity and effective temperature averaged over the stellar surface; the color-magnitude spread due to gravity darkening and random inclination angles is not shown. It is also important to mention that the behaviour between MIST and other stellar evolution models that include rotation such as Geneva \citep{ekstrom12} rotating models, may differ considerably due to the different assumptions for the convective and rotational mixing efficiencies (Choi et al. 2016, Gossage et al. 2018). In MIST, the effect of rotational mixing is relatively low in comparison with e.g.~Geneva and gravity darkening is the dominant effect, \textit{pushing} the MSTO region of the isochrone to cooler, redder areas of the CMD\footnote{We refer the reader to \citet{choi16, dotter16, choi18}, and \citet{gossage18} for a detailed description of the assumptions and limitations of MESA models, as well as comparisons with other stellar evolution models available}. However, we would like to add that we have also performed visual fits to the data with both rotating and non-rotating Geneva isochrones, finding consistent results to the ones presented here.} 

In either case (with or without stellar rotation), it appears that the color-magnitude spread of turn off stars in NGC~2509, while very narrow, still gives some uncertainty to the inferred cluster age, varying over about 75 Myr from roughly $\log(\rm{Age}) = 8.89$ to $8.95$. As mentioned before, this eMSTO spread is much smaller than expected (should be near 160 Myr according to the trend presented in \citet{niederhofer15}) for a cluster of this age.

\begin{figure}
\centerline{\includegraphics[scale=0.3]{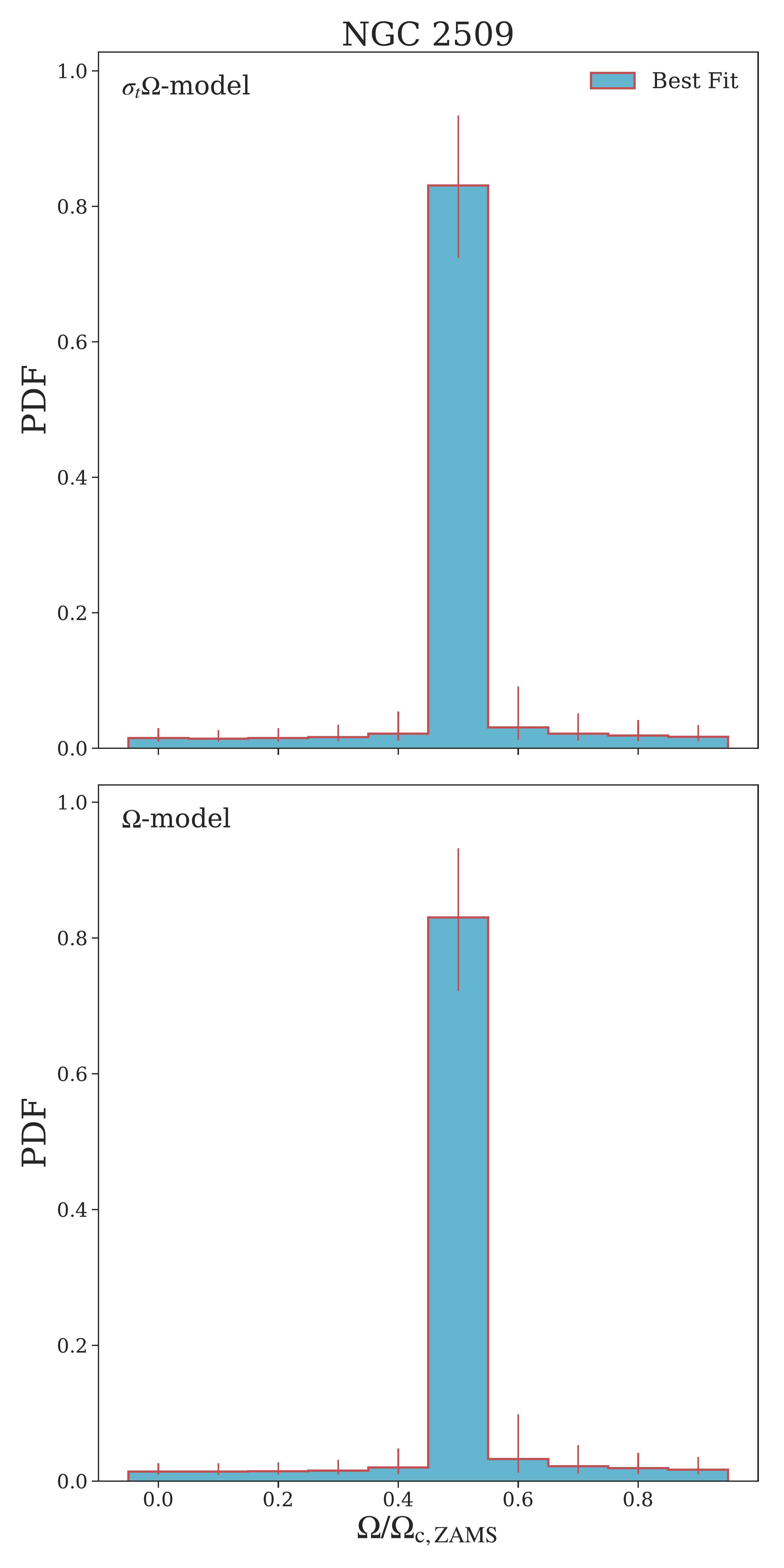}}
\caption{The resulting \vvc~distribution when employing the fit procedure of Gossage et al. (2019) for the $\sigma_t  \Omega$ model (which includes an age spread and a distribution in rotation rates, top panel) and the $\Omega$ model (only a distribution in rotation rates) in the bottom panel.  In both cases we find an extremely peaked distribution of the rotation rates.  In all cases we assumed a random orientation of the stellar rotation rates with respect to the line of sight.}
\label{fig:bestfitvcc}
\end{figure}

{While the {visual fit of isochrones} suggests that this data is best described with moderate stellar rotation of all stars in the cluster, the real constrains this qualitative analysis can put on the rotation distribution of the system are very limited. Thus, we followed it up with a dedicated statistical analysis. For this, we put our NGC~2509 data through the same simulation framework as in \citet{gossage18,gossage19}, where synthetic stellar populations are created with MATCH \citep{dolphin1997,dolphin02}, based on MIST stellar models, and are fitted to the data returning the best-fitting parameters for the observed cluster (see Section 3 of \citealp{gossage18} for a detailed description of the methodology used in this study). Gravity darkening is incorporated in these synthetic clusters considering a random distribution of inclination angles with respect to the line of sight. The fit between data and synthetic model cluster is then carried out via Hess diagrams and Poisson likelihood statistics, which are combined to find the overall likelihood of the data-model comparison between the models considered.} 

{For our analysis, we build three different models that we refer to as $\sigma_t $, the $\sigma_t \Omega$, and the $\Omega$ models, which allow for only age-spread, age-spread combined with stellar rotation rate distribution within the cluster, and only stellar rotation rate distribution, respectively. We consider ages in the range of $\log(\rm{Age}) = 7.60$ to $9.80$, and rotation rates from \vvc$= 0.0$ to $0.9$. We then fit the Gaia data of NGC~2509, non-parametrically, finding that the latter two models provide a significantly better overall likelihood than the $\sigma_t $~ model. These two best fitting models then score about the same probability and consistently find an extremely narrow distribution of stellar rotation rates centred at \vvc $=0.5$, with the $\Omega$-model finding a best fit age for the cluster at $\log(\rm{Age}) = 8.89$, and the $\sigma_t \Omega$-model combining the peaked \vvc$=0.5$ distribution with an age spread of $13~\rm{Myr}$, centred at $\log(\rm{Age}) = 8.90$, which is also consistent with no age spread whatsoever. Figure \ref{fig:bestfitvcc} shows the "weight" on the y-axis and \vvc~on the x-axis for these two best fit models, with the $\sigma_t \Omega$ and $\Omega$ models shown in the top and bottom panels, respectively. So it seems that, within the constraints of our simulation framework, the cause of the narrow MSTO in NGC~2509 is a very narrow distribution of stellar rotation rates, i.e.~all stars seem to be rotating at the same rate of near \vvc=0.5 in this cluster. Finally, we note that these simulations cannot provide tight constrainst on the orientation of the rotating stars aside from saying that it is consistent with a random distribution. However, we can say that, while there is reason to think that the stellar spins may be aligned in some clusters (see \citet{kamann19}), such an alignment could not reproduce the observed narrow MSTO in NGC~2509 by itself, i.e., a narrow \vvc~distribution is always required, according to our results.} 

{We would like to emphasise that these simulations are here presented simply as a plausible explanation of the observed morphology of the MSTO in NGC~2509 and that the only way to confirm if stellar rotation is indeed causing this particular feature is to obtain $Vsini$ measurements of MSTO stars in this cluster.}

%======================
\subsection{Binaries}\label{sec:binaries}
\begin{figure}
\centerline{\includegraphics[scale=0.35]{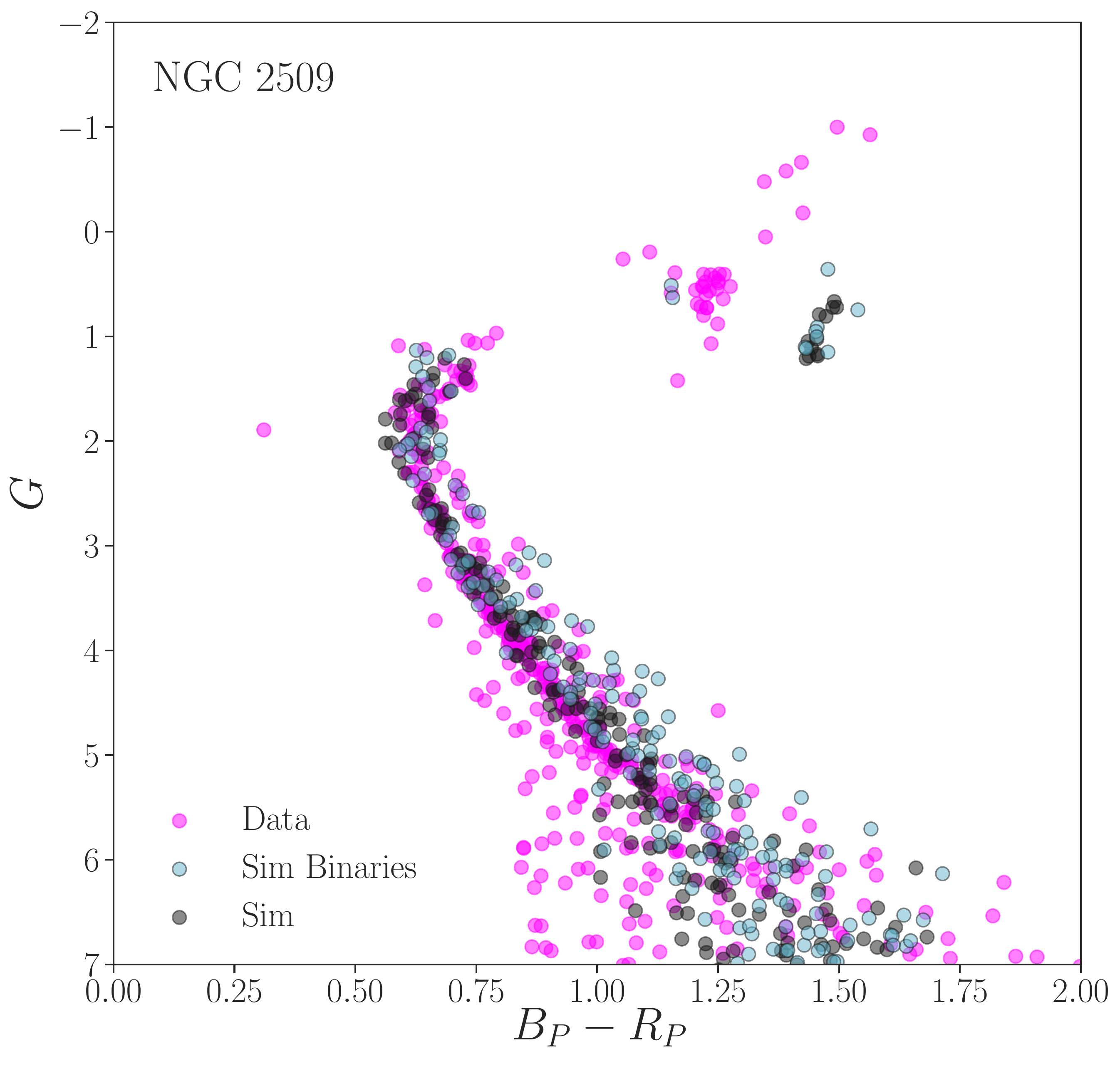}}
\caption{CMD of both NGC~2509 data and synthetic cluster created based on our best fit parameters ($\Omega$-model, see text). Gaia data is shown as magenta-filled dots, while synthetic data is shown in grey and blue-filled dots, with blue dots highlighting the binaries in our synthetic population.}
\label{fig:binaries}
\end{figure}

Finally, we looked at the binary fraction in NGC~2509 which could also have an impact on the spread (or lack thereof) in the MSTO.  We define the binary fraction as 
\begin{equation}
f_{\rm{bin}}=N_{\rm{bin}}/N_{\rm{tot}},
\end{equation}
where $N_{\rm{bin}}$ is the number of members in the binary sequence along our specified MS region of each cluster, and $N_{\rm{tot}}$ is the total number of members in that same region. We define a star as member of the binary sequence when its distance to the line B in the left panel of Figure~\ref{fig:distributions} is larger than $0.375$ (once all peaks have been aligned with NGC~2509's one).  This is approximately the inflexion or minimum point (depending on the cluster) between the two peaks of the MS distribution, separating the main sequence and binary sequence (gray-dashed vertical line in lower-right panel of Fig.~\ref{fig:distributions}). We use the same value for all clusters which might not be exact but it is a reasonable approximation looking at our figure. The binary fractions we find following this criterion are listed in Table~\ref{tab:parameters}. But we need to keep in mind that these fractions are not complete, since we are only sensitive to nearly equal mass binaries in these bands. 

In order to correct for the unseen fraction of binaries in NGC~2509 we performed simulations creating synthetic clusters and comparing the synthetic CMD with the observed one. We simulated NGC~2509 assuming $M = 3100$~\Msun, $f_{\rm{bin}} = 17\%$, $\log(\rm{Age}) = 8.89$, \vvc $= 0.5$, and a {random distribution of inclination angles}, as our analysis of the rotation distribution above suggests. We note that binaries are modelled in our code as having a flat mass fraction distribution (0.0 to 1.0), with no interaction between the stars.  We then increased the binary fraction in our simulated cluster until the total number of stars in the MS region of the simulated cluster, where our observations should be complete, matches the observed number of stars in NGC~2509 in that range. We find that the fraction of binaries should be about $50\%$, which is quite a high fraction for a cluster this age. Figure~\ref{fig:binaries} shows the CMDs of both synthetic and observed NGC~2509. The remarkable thing about this simulation is that it shows how, despite the high fraction of binaries included, these don't seem to have a broadening effect on the MSTO.

%%%%%%%%%%%%%%%%%%%%1%%%%%%%%%%%%%%%%
\section{Conclusions}
\label{sec:conclusions}

In this study we present Gaia photometry of several galactic open clusters with ages around $900~\rm{Myr}$ and compare their MSTO morphologies.  We do this by directly comparing the morphology of the MSTO region of the CMD in all clusters, instead of estimating a spread by fitting isochrones.  We focus on the open cluster NGC~2509 which lacks an extended MSTO while all other open clusters we could find at this age do show this feature.  Through our results and analysis we arrive to the following conclusions:

\begin{itemize}

\item{NGC~2509 does not show a spread in its MSTO while all other clusters at similar ages in our sample do. } 

\item{The differences in morphology between NGC~2509 and other clusters are statistically significant and not likely to be caused by effects such as e.g.~differential extinction. Stellar density effects on the rotation distribution of the stars in the cluster, which also could affect the width of the MSTO, are also ruled out as NGC~2509 does not seem to be any different than the others in that respect.}

\item{Stellar evolution models that include the effect of rotation are able to reproduce the particular characteristics of NGC~2509's MSTO much better than non-rotating models can. In particular, the models best fit the data when assuming a narrow range of rotational velocities centred on \vvc$ = 0.5$.  This is true independently of the age and rotational inclination angle in the ranges we've tested {within both MIST and Geneva frameworks}.}

\item{The binary fraction we measure from observations in NGC~2509 is of $\sim17\%$ but we are limited in that the Gaia bands we use are only sensitive to similar-mass binaries. We simulate the cluster based on the properties of the rotation distribution we find as a best fit in our analysis and find a total binary fraction of $\sim50\%$. This high fraction however does not seem to have an effect on the width of the MSTO, and so we rule out the influence of binaries as a possible cause of the big differences we see between the morphology of NGC~2509's CMD and that of the rest of clusters in our sample.}

\end{itemize}

{In light of these results, it is clear that NGC~2509 is a very special cluster whose unique properties could potentially reveal a great deal regarding if and how stellar rotation affects the morphology of a cluster's CMD, and, as such, it deserves further investigation. In particular, it is pressing to obtain data on the stellar rotational distribution to confirm its peaked nature.  Additionally, we will be exploring our Gaia DR2 open cluster database to look for other clusters with anomalous MSTOs.} 

\section{Acknowledgements}

MdJO, SK, CU and NB gratefully acknowledge funding from a European Research Council consolidator grant (ERC-CoG-646928-Multi-Pop).  NB is a Royal Society University Research Fellow. Support for this work was also provided by NASA through Hubble Fellowship grant HST-HF2-51387.001-A awarded by the Space Telescope Science Institute, which is operated by the Association of Universities for Research in Astronomy, Inc., for NASA, under contract NAS5-26555.

\bibliographystyle{mn2e}
\bibliography{./mjsbib.bib}

\begin{thebibliography}{26}
\expandafter\ifx\csname natexlab\endcsname\relax\def\natexlab#1{#1}\fi

\bibitem[{{Bastian} \& {de Mink}(2009)}]{bastiandemink09}
{Bastian} N., {de Mink} S.~E., 2009, \mnras, 398, L11

\bibitem[{{Bastian} {et~al}\mbox{.}(2018){Bastian}, {Kamann}, {Cabrera-Ziri},
  {Georgy}, {Ekstr{\"o}m}, {Charbonnel}, {de Juan Ovelar}, \&
  {Usher}}]{bastian18}
{Bastian} N. {et~al.}, 2018, \mnras, 480, 3739

\bibitem[{{Beasor} {et~al}\mbox{.}(2019){Beasor}, {Davies}, {Smith}, \&
  {Bastian}}]{beasor19}
{Beasor} E.~R. {et~al.}, 2019, \mnras, 486, 266

\bibitem[{{Brandt} \& {Huang}(2015)}]{brandthuang15b}
{Brandt} T.~D., {Huang} C.~X., 2015, \apj, 807, 25

\bibitem[{{Choi} {et~al}\mbox{.}(2018){Choi}, {Conroy}, {Ting}, {Cargile},
  {Dotter}, \& {Johnson}}]{choi18}
{Choi} J. {et~al.}, 2018, \apj, 863, 65

\bibitem[{{Choi} {et~al}\mbox{.}(2016){Choi}, {Dotter}, {Conroy}, {Cantiello},
  {Paxton}, \& {Johnson}}]{choi16}
{Choi} J. {et~al.}, 2016, \apj, 823, 102

\bibitem[{{Cordoni} {et~al}\mbox{.}(2018){Cordoni}, {Milone}, {Marino}, {Di
  Criscienzo}, {D'Antona}, {Dotter}, {Lagioia}, \& {Tailo}}]{cordoni18}
{Cordoni} G. {et~al.}, 2018, \apj, 869, 139

\bibitem[{{Dolphin}(1997)}]{dolphin1997}
{Dolphin} A., 1997, \na, 2, 397

\bibitem[{{Dolphin}(2002)}]{dolphin02}
{Dolphin} A.~E., 2002, \mnras, 332, 91

\bibitem[{{Dotter}(2016)}]{dotter16}
{Dotter} A., 2016, \apjs, 222, 8

\bibitem[{{Dupree} {et~al}\mbox{.}(2017){Dupree}, {Dotter}, {Johnson},
  {Marino}, {Milone}, {Bailey}, {Crane}, {Mateo}, \& {Olszewski}}]{dupree17}
{Dupree} A.~K. {et~al.}, 2017, \apjl, 846, L1

\bibitem[{{Ekstr{\"o}m} {et~al}\mbox{.}(2012){Ekstr{\"o}m}, {Georgy},
  {Eggenberger}, {Meynet}, {Mowlavi}, {Wyttenbach}, {Granada}, {Decressin},
  {Hirschi}, {Frischknecht}, {Charbonnel}, \& {Maeder}}]{ekstrom12}
{Ekstr{\"o}m} S. {et~al.}, 2012, \aap, 537, A146

\bibitem[{{Gossage} {et~al}\mbox{.}(2019){Gossage}, {Conroy}, {Dotter},
  {Cabrera-Ziri}, {Dolphin}, {Bastian}, {Dalcanton}, {Goudfrooij}, {Johnson},
  {Williams}, {Rosenfield}, {Kalirai}, \& {Fouesneau}}]{gossage19}
{Gossage} S. {et~al.}, 2019, arXiv e-prints, arXiv:1907.11251

\bibitem[{{Gossage} {et~al}\mbox{.}(2018){Gossage}, {Conroy}, {Dotter}, {Choi},
  {Rosenfield}, {Cargile}, \& {Dolphin}}]{gossage18}
{Gossage} S. {et~al.}, 2018, \apj, 863, 67

\bibitem[{{Kamann} {et~al}\mbox{.}(2018){Kamann}, {Bastian}, {Husser},
  {Martocchia}, {Usher}, {den Brok}, {Dreizler}, {Kelz}, {Krajnovi{\'c}}, \&
  {Richard}}]{kamann18b}
{Kamann} S. {et~al.}, 2018, \mnras, 480, 1689

\bibitem[{{Kamann} {et~al}\mbox{.}(2019){Kamann}, {Bastian}, {Gieles},
  {Balbinot}, \& {H{\'e}nault-Brunet}}]{kamann19}
{Kamann} S. {et~al.}, 2019, \mnras, 483, 2197

\bibitem[{{Kamann} {et~al}\mbox{.}(2014){Kamann}, {Wisotzki}, {Roth},
  {Gerssen}, {Husser}, {Sandin}, \& {Weilbacher}}]{kamann14}
{Kamann} S. {et~al.}, 2014, \aap, 566, A58

\bibitem[{{Li} {et~al}\mbox{.}(2017){Li}, {de Grijs}, {Deng}, \&
  {Milone}}]{li17}
{Li} C. {et~al.}, 2017, \apj, 844, 119

\bibitem[{{Lindegren} {et~al}\mbox{.}(2018){Lindegren}, {Hern{\'a}ndez},
  {Bombrun}, {Klioner}, {Bastian}, {Ramos-Lerate}, {de Torres},
  {Steidelm{\"u}ller}, {Stephenson}, {Hobbs}, {Lammers}, {Biermann}, {Geyer},
  {Hilger}, {Michalik}, {Stampa}, {McMillan}, {Casta{\~n}eda}, {Clotet},
  {Comoretto}, {Davidson}, {Fabricius}, {Gracia}, {Hambly}, {Hutton}, {Mora},
  {Portell}, {van Leeuwen}, {Abbas}, {Abreu}, {Altmann}, {Andrei}, {Anglada},
  {Balaguer-N{\'u}{\~n}ez}, {Barache}, {Becciani}, {Bertone}, {Bianchi},
  {Bouquillon}, {Bourda}, {Br{\"u}semeister}, {Bucciarelli}, {Busonero},
  {Buzzi}, {Cancelliere}, {Carlucci}, {Charlot}, {Cheek}, {Crosta}, {Crowley},
  {de Bruijne}, {de Felice}, {Drimmel}, {Esquej}, {Fienga}, {Fraile}, {Gai},
  {Garralda}, {Gonz{\'a}lez-Vidal}, {Guerra}, {Hauser}, {Hofmann}, {Holl},
  {Jordan}, {Lattanzi}, {Lenhardt}, {Liao}, {Licata}, {Lister}, {L{\"o}ffler},
  {Marchant}, {Martin-Fleitas}, {Messineo}, {Mignard}, {Morbidelli}, {Poggio},
  {Riva}, {Rowell}, {Salguero}, {Sarasso}, {Sciacca}, {Siddiqui}, {Smart},
  {Spagna}, {Steele}, {Taris}, {Torra}, {van Elteren}, {van Reeven}, \&
  {Vecchiato}}]{lindegren18}
{Lindegren} L. {et~al.}, 2018, \aap, 616, A2

\bibitem[{{Mackey} \& {Broby Nielsen}(2007)}]{mackeynielsen07}
{Mackey} A.~D., {Broby Nielsen} P., 2007, \mnras, 379, 151

\bibitem[{{Maeder}(2009)}]{maeder09}
{Maeder} A., 2009, {Physics, Formation and Evolution of Rotating Stars}

\bibitem[{{Marino} {et~al}\mbox{.}(2018){Marino}, {Milone}, {Casagrande},
  {Przybilla}, {Balaguer-N{\'u}{\~n}ez}, {Di Criscienzo}, {Serenelli}, \&
  {Vilardell}}]{marino18b}
{Marino} A.~F. {et~al.}, 2018, \apjl, 863, L33

\bibitem[{{Martocchia} {et~al}\mbox{.}(2018){Martocchia}, {Niederhofer},
  {Dalessandro}, {Bastian}, {Kacharov}, {Usher}, {Cabrera-Ziri}, {Lardo},
  {Cassisi}, {Geisler}, {Hilker}, {Hollyhead}, {Kozhurina-Platais}, {Larsen},
  {Mackey}, {Mucciarelli}, {Platais}, \& {Salaris}}]{martocchia18}
{Martocchia} S. {et~al.}, 2018, \mnras, 477, 4696

\bibitem[{{Niederhofer} {et~al}\mbox{.}(2015){Niederhofer}, {Georgy},
  {Bastian}, \& {Ekstr{\"o}m}}]{niederhofer15}
{Niederhofer} F. {et~al.}, 2015, \mnras, 453, 2070

\bibitem[{{Piatti} \& {Bastian}(2016)}]{piattibastian16}
{Piatti} A.~E., {Bastian} N., 2016, \aap, 590, A50

\bibitem[{{von Zeipel}(1924)}]{vonzeipel1924}
{von Zeipel} H., 1924, \mnras, 84, 665

\end{thebibliography}

\end{document}